\documentclass[aps,showpacs,twocolumn,superscriptaddress]{revtex4}

\usepackage{graphicx}
\usepackage{dcolumn}

\begin{document}

\title{Light $\Lambda\Lambda$ Hypernuclei and the Onset of Stability 
for $\Lambda\Xi$ Hypernuclei}

\author{I.N.~Filikhin}
\affiliation{Racah Institute of Physics, The Hebrew University, 
Jerusalem 91904, Israel\vspace*{1ex}}

\author{A.~Gal}
\affiliation{Racah Institute of Physics, The Hebrew University, 
Jerusalem 91904, Israel\vspace*{1ex}}


\begin{abstract}
\rule{0ex}{3ex}

New Faddeev-Yakubovsky calculations for light $\Lambda\Lambda$ 
hypernuclei are presented in order to assess the self consistency 
of the $\Lambda\Lambda$-hypernuclear binding-energy world data 
and the implied strength of the $\Lambda\Lambda$ interaction, 
in the wake of recent experimental reports on 
$_{\Lambda\Lambda}^{~4}$H and $_{\Lambda\Lambda}^{~6}$He. 
Using Gaussian soft-core simulations of Nijmegen one-boson-exchange 
model interactions, the Nijmegen soft-core model NSC97 simulations  
are found close to reproducing the recently reported binding 
energy of $_{\Lambda\Lambda}^{~6}$He, but not those of other species. 
For stranger systems, Faddeev calculations of light $\Lambda\Xi$ 
hypernuclei, using a simulation of the strongly attractive 
$\Lambda\Xi$ interactions due to the same model, suggest that 
$_{\Lambda\Xi}^{~6}$He marks the onset of nuclear stability 
for $\Xi$ hyperons. 
\end{abstract} 

\maketitle

\section{Introduction} 

Very little is known experimentally on doubly-strange hypernuclear 
systems, and virtually nothing about systems with higher strangeness 
content. Multistrange hadronic matter in finite systems and in bulk 
is predicted on general grounds to be stable, up to strangeness 
violating weak decays (\cite{SBG00} and references therein). 
Hyperons ($Y$) must contribute macroscopically to 
the composition of neutron-star matter (\cite{Gle01} and references 
therein). Over the years the Nijmegen group has constructed 
a number of one-boson-exchange (OBE) models for the baryon-baryon 
interaction, fitting the abundant scattering and 
bound-state $NN$ data plus the scarce and poorly determined 
low-energy $YN$ data using SU(3)-flavor symmetry to relate 
baryon-baryon-meson coupling constants and phenomenological 
short-distances hard or soft cores (\cite{Rij01} and references 
therein). Data on multistrange systems could help distinguishing 
between these models. The recently reported events from AGS 
experiment E906 suggest production of light $\Lambda\Lambda$ 
hypernuclei \cite{Ahn01}, perhaps as light even as 
$_{\Lambda\Lambda}^{~4}$H, in the $(K^-,K^+)$ reaction on $^9$Be. 
If $^{~4}_{\Lambda\Lambda}$H is confirmed in a future extension 
of this experiment, this four-body system $pn \Lambda \Lambda$ 
would play as a fundamental role for studying theoretically 
the $YY$ forces as $^{3}_{\Lambda}$H $(pn\Lambda)$ has played 
for studying theoretically the $YN$ forces \cite{MKG95}. 

Until recently only three $\Lambda\Lambda$ hypernuclear candidates 
fitted events seen in emulsion experiments \cite{Dan63,Pro66,Aok91}. 
The $\Lambda\Lambda$ binding energies deduced from these `old' 
events suggest a strongly attractive $\Lambda\Lambda$ interaction in 
the $^{1}S_0$ channel \cite{DMGD91}. This outlook might be 
changing substantially following the very recent report from the 
hybrid-emulsion KEK experiment E373 on a new event \cite{Tak01} 
uniquely interpreted as $_{\Lambda\Lambda}^{~6}$He, with binding 
energy considerably smaller than that reported for the older 
event \cite{Pro66}. 

In this Rapid Communication we report on new Faddeev-Yakubovsky 
calculations for light $\Lambda\Lambda$ hypernuclei, using generic 
$s$-wave $\Lambda\Lambda$ interaction potentials which simulate 
the low-energy $s$-wave scattering parameters produced by the 
Nijmegen OBE models. The purpose of these calculations is twofold: 
to check the self consistency of the data, particularly for 
$_{\Lambda\Lambda}^{~6}$He and $_{\Lambda\Lambda}^{10}$Be 
which are treated here as clusters of $\alpha$'s and $\Lambda$'s; 
and to find out which of the Nijmegen OBE models is the most 
appropriate one for describing these $\Lambda\Lambda$ hypernuclei. 

A novel piece of work concluding this report concerns 
multistrange hypernuclei consisting, in addition to 
$\Lambda$'s, also of a (doubly strange $S=-2$) $\Xi$ 
hyperon. Schaffner {\it et al.} \cite{Sch92} observed 
that $\Xi$ hyperons would become particle stable against 
the strong decay $\Xi N \rightarrow \Lambda \Lambda$ 
if a sufficient number of bound $\Lambda$'s Pauli-blocked 
this decay mode, highlighting 
$_{\Lambda\Lambda\Xi}^{~~~7}$He ($S=-4$) as the lightest 
system of its kind. Here we study the possibility 
of stabilizing a $\Xi$ hyperon in the isodoublet 
$_{\Lambda\Xi}^{~6}$H - $_{\Lambda\Xi}^{~6}$He ($S=-3$) 
hypernuclei due to the particularly strong $\Lambda\Xi$ 
attraction in the Nijmegen Soft Core NSC97 model \cite{RSY99}. 
This three-body $\alpha\Lambda\Xi$ system may provide the 
onset of $\Xi$ nuclear stability

\section{Methodology and input} 

In our calculations, the bound states of three- and four-body systems 
are obtained by solving the differential $s$-wave Faddeev-Yakubovsky 
equations \cite{MFa93}, using the cluster reduction method \cite{YFi95} 
in which the various channel wavefunctions are decomposed in terms 
of eigenfunctions of the Hamiltonians of the two- or three-particle 
subsystems. A fairly small number of terms, generally less than 10, 
is sufficient to generate a stable and precise numerical solution.
This method has been recently applied to $_{\Lambda}^9$Be and 
$_{\Lambda\Lambda}^{~6}$He in terms of three-cluster 
$\alpha\alpha\Lambda$ and $\alpha\Lambda\Lambda$ systems, 
respectively \cite{FYa00}.  

\begin{figure}[t] 
\centerline{\includegraphics[height=6.5cm]{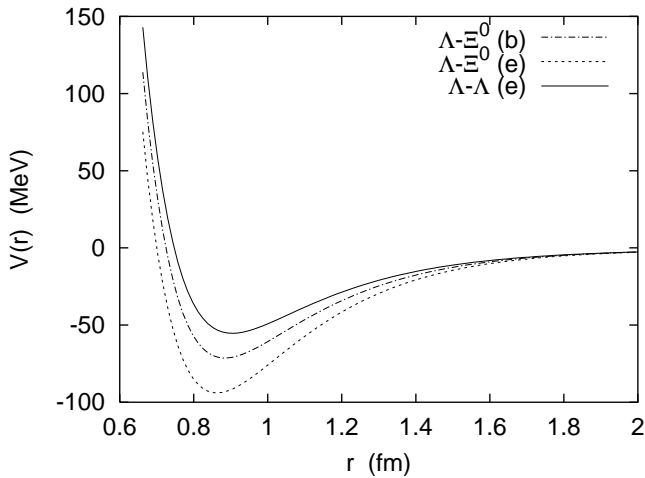}} 
\caption{Selected hyperon-hyperon potentials, simulating versions $b$ and 
$e$ of the NSC97 model interactions \cite{RSY99}.} 
\label{fig:pot-LY} 
\end{figure}

The hyperon-hyperon interaction potentials in the $^{1}S_0$ 
channel which are used as input to the above equations are of a 
three-range Gaussian form 
\begin{equation} 
\label{eq:HKM}
V_{YY'} = \sum_i^3 v_{YY'}^{(i)}(r)\exp(-r^2/\beta_i^2)\;\;, 
\end{equation} 
following the work of Hiyama {\it et al.} \cite{HKM97} where 
a $\Lambda\Lambda$ potential of this form was fitted to the Nijmegen 
model D (ND) hard-core interaction \cite {Nag75} assuming the same hard 
core for the $NN$ and $\Lambda\Lambda$ potentials in the $^{1}S_0$ channel. 
For other models we have renormalized the medium-range attractive component 
($i=2$) of this potential such that it yields values for the $s$-wave 
scattering length and for the effective range as close to the values produced 
by Nijmegen model interaction potentials for these low-energy parameters. 
Several $YY$ potentials fitted to the low-energy parameters of the 
soft-core NSC97 model \cite{RSY99} are shown in Fig. \ref{fig:pot-LY}. 
We note that the $\Lambda\Xi$ interaction is rather strong, considerably 
stronger within the same version of the model (here $e$) than the 
$\Lambda\Lambda$ interaction. The $\Lambda\Lambda$ interaction is fairly 
weak for all of the six versions ($a$)-($f$) of model NSC97.

The $\alpha\alpha$ short-range interaction, and the $\Lambda\alpha$ and 
$\Xi\alpha$ interactions, are given in terms of a two-range Gaussian 
(Isle) potential 
\begin{equation} 
\label{eq:Isle}
V_{Y\alpha}=V_{\rm rep}^{(Y)}\exp (-r^2/\beta_{\rm rep}^2)-
V_{\rm att}^{(Y)}\exp (-r^2/\beta_{\rm att}^2)\;\;.
\end{equation} 
Here the superscript $Y$ extends also for $\alpha$. 
For the $\alpha\alpha$ short-range potential we used the 
$s$-wave component of the Ali-Bodmer potential \cite{ABo66}.
A finite-size Coulomb potential was added. 
The $\Lambda\alpha$ potential, fitted to the binding energy 
$B_{\Lambda}(^5_{\Lambda}\rm He) = 3.12 \pm 0.02$ MeV \cite{DPn86}, 
was taken from Ref. \cite{KOA95}. For the $\Xi\alpha$ potential 
we assumed $V_{\rm rep}^{(\Xi)} = V_{\rm rep}^{(\Lambda)}$ 
while reducing the depth $V_{\rm att}^{(\Lambda)}$ to get 
$\Xi^0\alpha$ binding energy 2.09 MeV. This $B_{\Xi}$ value 
was obtained using a Woods-Saxon (WS) potential for $^4$He 
with a depth parameter scaled by the ratio of central densities 
with respect to a depth of $\sim$15 MeV in $^{11}$B, as suggested 
by studying the excitation spectrum in the ($K^-,K^+$) reaction 
on $^{12}$C \cite{Fuk98}.  

\section{Results and discussion} 

\subsection{$\Lambda\Lambda$ hypernuclei} 

We first applied, for a test, these $\alpha\alpha$ and 
$\Lambda\alpha$ potentials (\ref{eq:Isle}) in a three-body 
$s$-wave Faddeev calculation for the $\alpha\alpha\Lambda$ 
system. We will comment below on the restriction to $s$ waves. 
The calculated ground-state binding energy, 
$B_{\Lambda}(^9_{\Lambda}\rm Be) = 6.67$ MeV, is in excellent 
agreement with the measured value $6.71 \pm 0.04$ MeV \cite{DPn86} 
without need for renormalization \cite{HKM97} or for introducing 
three-body interactions \cite{BUC84}. We then applied these 
potentials in Faddeev-Yakubovsky calculations for several 
$\Lambda\Lambda$ hypernuclei, using $\Lambda\Lambda$ interactions 
generically of the form (\ref{eq:HKM}) which simulate some of the 
Nijmegen OBE interaction potentials. The results are stable 
against reasonable variations in the $\Lambda\Lambda$ potential 
shape, provided the underlying low-energy parameters are kept 
fixed. The ground-state $\Lambda\Lambda$ binding energies 
$B_{\Lambda\Lambda}$ obtained by solving the $s$-wave 
three-body $(\alpha\Lambda\Lambda)$ Faddeev equations for 
$^{~6}_{\Lambda\Lambda}$He and the $s$-wave four-body 
$(\alpha\alpha\Lambda\Lambda)$ Yakubovsky equations for 
$^{10}_{\Lambda\Lambda}$Be are given in Table \ref{tab:tabl3}. 
Using the ND-simulated $\Lambda\Lambda$ interaction our results 
may be compared with those of Ref. \cite{HKM97} which were not 
limited to the dominant $s$-wave channels. 
For $^{~6}_{\Lambda\Lambda}$He, and with similar $\Lambda\alpha$ 
potentials, the inclusion of higher ($d$) partial waves amounts 
to additional 0.2 MeV binding. For $^{10}_{\Lambda\Lambda}$Be 
the effect of the higher partial waves is largely compensated 
by keeping $B_{\Lambda}(^9_{\Lambda}$Be) at its experimental 
value, whether or not including $d$ waves. This was also the 
practice in Ref. \cite{HKM97}; the comparison in 
Table \ref{tab:tabl3} suggests an effect of order 0.5 MeV, 
which is similar to the effect of model dependence due to using 
different underlying $\Lambda N$ interaction potentials in that 
work. Focussing on our own calculations, Table \ref{tab:tabl3} 
shows that the strongest $\Lambda\Lambda$ binding is provided 
by the simulation of the very recent extended soft core (ESC00) 
model \cite{Rij01} which was in fact motivated by the relatively 
large $B_{\Lambda\Lambda}$ value for the 
$_{\Lambda\Lambda}^{~6}$He `old' event \cite{Pro66}. 
A significantly smaller $B_{\Lambda\Lambda}$ value is obtained for 
our simulation of model ND which, however, reproduces well the 
$B_{\Lambda\Lambda}$ value reported for $_{\Lambda\Lambda}^{10}$Be 
\cite{Dan63}. Down the list, the simulation of the NSC97 model 
gives yet smaller $B_{\Lambda\Lambda}$ values, which for 
$_{\Lambda\Lambda}^{~6}$He are close to the very recent 
experimental report \cite{Tak01} almost independently of 
which version of the model is used.

\begin{table}
\caption{Calculated ground-state binding energies  
($B_{\Lambda\Lambda}$ in MeV with respect to the nuclear core).} 
\label{tab:tabl3} 
\begin{tabular}{ccc} 
\hline \hline  
 Model & $^{~6}_{\Lambda\Lambda}$He & $^{10}_{\Lambda\Lambda}$Be\\ 
 \hline 
 ESC00 & 10.7 & 19.5 \\
  ND   & 9.10 & 17.8 \\
 NSC97$e$ & 6.82 & 15.5 \\
 NSC97$b$ & 6.60 & 15.3 \\ 
 $V_{\Lambda\Lambda}=0$ & 6.27 & 14.9 \\ \hline 
 \cite{HKM97}(ND) & 9.34 & 17.24\\  
 \hline
 exp. & 10.9$\pm$0.6 \cite{Pro66} & 17.7$\pm$0.4 \cite{Dan63} \\
 & 7.25$\pm$0.19$^{+0.18}_{-0.11}$ \cite{Tak01} 
 & (14.6$\pm$0.4) \footnote{assuming 
 ~$^{10}_{\Lambda\Lambda}$Be ~$\to~ \pi^- ~+~ p ~+~ ^9_\Lambda$Be*} \\ 
\hline \hline 
\end{tabular}
\end{table}

\begin{figure}[t] 
\centerline{\includegraphics[height=6.5cm]{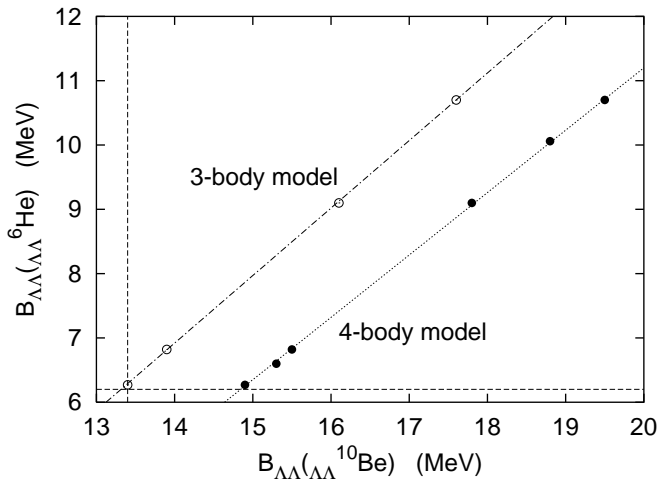}} 
\caption{Calculated binding energies ($B_{\Lambda\Lambda}$ in MeV) 
for $^{~6}_{\Lambda\Lambda}$He in a three-body $\alpha\Lambda\Lambda$ 
model, and for $_{\Lambda\Lambda}^{10}$Be in a four-body 
$\alpha\alpha\Lambda\Lambda$ model and in a three-body 
~$^8$Be $\Lambda \Lambda$ model. The origin of the dashed axes 
corresponds to $\Delta B_{\Lambda\Lambda} = 0$.} 
\label{fig:e6e10} 
\end{figure}

Early cluster calculations \cite{BUC84,WTB86} noted 
that the calculated $B_{\Lambda\Lambda}$ values for 
$^{~6}_{\Lambda\Lambda}$He and for $^{10}_{\Lambda\Lambda}$Be are 
correlated nearly linearly with each other, such that the two 
events reported in the 60's could not be reproduced simultaneously.  
Our calculations also produce such a correlation, as demonstrated 
in Fig. \ref{fig:e6e10} by the solid circles along the dotted line. 
This line precludes any joint theoretical framework in terms of 
two-body interactions alone for the $^{~6}_{\Lambda\Lambda}$He 
and $^{10}_{\Lambda\Lambda}$Be experimental candidates listed in 
Table \ref{tab:tabl3}. For $V_{\Lambda\Lambda}$ = 0, the 
lower-left point on the dotted line corresponds to approximately 
zero incremental binding energy $\Delta B_{\Lambda\Lambda}$ 
for $^{~6}_{\Lambda\Lambda}$He, where 
\begin{equation} 
\label{eq:delB} 
\Delta B_{\Lambda\Lambda} (^{~A}_{\Lambda \Lambda}Z) 
= B_{\Lambda\Lambda} (^{~A}_{\Lambda \Lambda}Z) 
- 2B_{\Lambda} (^{(A-1)}_{~~\Lambda}Z)\;\;. 
\end{equation} 
This is easy to understand owing to the rigidity of the $\alpha$ core. 
However, the corresponding $\Delta B_{\Lambda\Lambda}$ value for 
$^{10}_{\Lambda\Lambda}$Be is fairly substantial, about 1.5 MeV, 
reflecting a basic difference between the four-body 
$\alpha\alpha\Lambda\Lambda$ calculation and any three-body 
approximation in terms of a nuclear core and two $\Lambda$'s as in 
$^{~6}_{\Lambda\Lambda}$He. To demonstrate this point we show 
by the open circles along the dot-dash line 
in Fig.~\ref{fig:e6e10} the results of a three-body calculation for 
$^{10}_{\Lambda\Lambda}$Be in which the $^{8}$Be core is not 
assigned an $\alpha\alpha$ structure. In this calculation,  
the geometry and depth of the $\Lambda-^{8}$Be WS potential 
were fitted to reproduce (i) the measured 
$B_{\Lambda}(^{9}_{\Lambda}$Be) value and (ii) the r.m.s. 
distance between the $\Lambda$ and the c.m. of the two 
$\alpha$'s as obtained in the $\alpha\alpha\Lambda$ model 
calculation for $^{9}_{\Lambda}$Be. This three-body 
$^{8}$Be$~\Lambda\Lambda$ calculation gives about 1.5 MeV less 
binding for $^{10}_{\Lambda\Lambda}$Be than the four-body 
calculation does. The difference is due to 
the $\alpha\alpha$ correlations which are absent in the 
three-body calculation, and which are built in within the 
Yakubovsky equations of the four-body calculation. 
The other calculations mentioned above \cite{BUC84,WTB86} found 
smaller values, not exceeding 0.5 MeV, for the binding-energy 
gain due to having a four-body calculation for 
$^{10}_{\Lambda\Lambda}$Be. An obvious merit of our four-body 
Faddeev-Yakubovsky calculation is that it automatically accounts 
for {\it all} possible rearrangement channels in the 
$\alpha\alpha\Lambda\Lambda$ system. In particular, by breaking 
up $^8$Be into two $\alpha$'s in the four-body calculation, 
substantial attraction is gained due to several additional bound 
subsystems such as the $^{~6}_{\Lambda\Lambda}$He - $\alpha$ and 
$^{5}_{\Lambda}$He - $^{5}_{\Lambda}$He clusters which almost 
saturate the corresponding rearrangement channels 
($\alpha\Lambda\Lambda$) - $\alpha$ and 
($\alpha\Lambda$) - ($\alpha\Lambda$), respectively. 

Our calculations confirm, if not aggravate, the incompatibility 
of the `old' experimental determination of the binding energy of 
$^{~6}_{\Lambda\Lambda}$He \cite{Pro66} with that of 
$^{10}_{\Lambda \Lambda}$Be \cite{Dan63}. 
The `new' experimental determination of the binding 
energy of $^{~6}_{\Lambda \Lambda}$He \cite{Tak01} is found 
to be still incompatible with that of $^{10}_{\Lambda \Lambda}$Be, 
even if an unobserved $\gamma$ deexcitation involving either 
$^{10}_{\Lambda \Lambda}$Be$^{*}$ or $^{9}_{\Lambda}$Be$^{*}$ 
is allowed for; one of these possibilities, involving 
$^{9}_{\Lambda}$Be$^*$ at 3.1 MeV \cite{May83}, is recorded in 
Table \ref{tab:tabl3}. Since no particle-stable excited states 
are possible for $^{~6}_{\Lambda \Lambda}$He or for its $\Lambda$ 
hypernuclear core $^{5}_{\Lambda}$He, and since 
$^{~6}_{\Lambda \Lambda}$He is also ideally suited for three-body 
cluster calculations such as the $s$-wave Faddeev equations here 
solved for the $\alpha \Lambda \Lambda$ system, we continue by 
discussing the implications of accepting the E373 KEK experiment 
\cite{Tak01} determination of $\Delta B_{\Lambda\Lambda}$~$\sim 1$~MeV 
for $^{~6}_{\Lambda \Lambda}$He. 
We have shown that model NSC97 is the only one capable of getting close 
to this `new' binding-energy value, short by about 0.5 MeV. In fact, 
we estimate the theoretical uncertainty of our Faddeev calculation for 
$^{~6}_{\Lambda \Lambda}$He as bounded by 0.5 MeV, and such that a more 
precisely calculated binding energy would be {\it larger} by a fraction 
of this bound, at most, than the $B_{\Lambda \Lambda}$ values shown in 
Table \ref{tab:tabl3}. Taking into account such possible corrections 
would bring our calculated $B_{\Lambda \Lambda}$ values to within the 
error bars of the reported $B_{\Lambda \Lambda}$ value. There are two 
possible origins for this theoretical uncertainty, one which was already 
mentioned above is the restriction to $s$-waves in the partial-wave 
expansion of the Faddeev equations; the other one is ignoring the 
off-diagonal $\Lambda\Lambda - \Xi N$ interaction which admixes $\Xi$ 
components into the $^{~6}_{\Lambda \Lambda}$He wavefunction. 
A recent work \cite{YNa00} using two $YN$ and $YY$ models finds 
an increase of 0.1 to 0.4 MeV in the calculated 
$B_{\Lambda \Lambda}(^{~6}_{\Lambda\Lambda}$He) value due to a 
0.1 to 0.3\% (probability) $\Xi$ component, respectively. 

\subsection{$\Lambda\Xi$ hypernuclei} 

\begin{figure}[t] 
\centerline{\includegraphics[height=6.8cm]{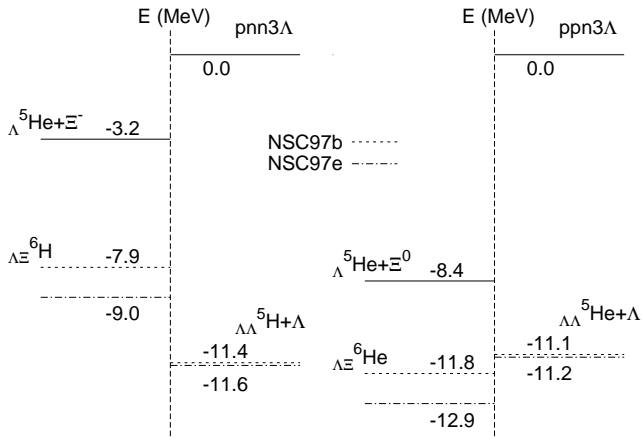}}
\caption{Calculated level scheme of $_{\Lambda\Xi}^{~6}$H and 
$_{\Lambda\Xi}^{~6}$He hypernuclei.} 
\label{fig:LXi} 
\end{figure}

If model NSC97 indeed provides for a valid extrapolation from fits 
to $NN$ and $YN$ data, and recalling the strongly attractive 
$^{1}S_{0}$ $\Lambda\Xi$ potentials in Fig. \ref{fig:pot-LY} 
simulating the NSC97 model, it is tempting to check for stability 
of $A=6, S=-3$ systems obtained from $^{~6}_{\Lambda\Lambda}$He by 
replacing a $\Lambda$ by a $\Xi$ hyperon. The results of a Faddeev 
calculation for the isodoublet hypernuclei $^{~6}_{\Lambda\Xi}$H and 
$^{~6}_{\Lambda\Xi}$He, considered as $\alpha\Lambda\Xi^-$ and 
$\alpha\Lambda\Xi^0$ three-body systems respectively, are shown 
in Fig. \ref{fig:LXi}, including the location of the lowest 
particle-stability thresholds, due to $\Lambda$ emission into 
$^{~5}_{\Lambda\Lambda}$H and $^{~5}_{\Lambda\Lambda}$He, respectively. 
These $A = 5$ isodoublet $\Lambda\Lambda$ hypernuclei, considered as 
three-cluster systems $^3$H$~\Lambda\Lambda$ and $^3$He~$\Lambda\Lambda$ 
respectively, are found to be particle stable for {\it all} the 
$\Lambda\Lambda$ attractive potentials used in the present calculation. 
Figure \ref{fig:LXi} demonstrates that $^{~6}_{\Lambda\Xi}$He is 
particle-stable for potentials simulating the NSC97 model. 
The mirror hypernucleus $_{\Lambda\Xi}^{~6}$H is unstable because the 
$\Xi^-$ hyperon is heavier by 6.5 MeV than $\Xi^0$. Our prediction 
for the stability of $_{\Lambda\Xi}^{~6}$He would hold valid, 
particularly for potentials simulating model NSC97$e$ (and also $f$), 
even if the binding energy of $_{\Lambda\Lambda}^{~5}$He is increased 
by a fraction of an MeV to scale it with the recently reported 
$_{\Lambda\Lambda}^{~6}$He binding energy \cite{Tak01}. 
However, if the $\Xi\alpha$ WS potential depth is set equal to that 
for $\Xi$ in $^{11}$B \cite{Fuk98}, $^{~6}_{\Lambda\Xi}$He would become 
unstable by a fraction of an MeV in version $e$. Lack of direct 
experimental evidence on $\Xi$ interactions in or around $^4$He prevents 
us from reaching a more definitive conclusion on this issue. 

\section{Conclusions} 

In summary, we have shown that $s$-wave simulations of the 
OBE Nijmegen model NSC97, versions $e$ and $f$ of which have 
been shown recently to agree quantitatively with light single 
$\Lambda$ hypernuclei \cite{AHS00}, 
are capable of reproducing the recently reported binding energy 
of $_{\Lambda\Lambda}^{~6}$He, but are incapable of reproducing 
previously reported $\Lambda\Lambda$ binding energies. 
This inconsistency, for a wide class of $\Lambda\Lambda$ 
potentials, was demonstrated on firm grounds by doing the 
first ever Faddeev-Yakubovsky calculation of 
$_{\Lambda\Lambda}^{10}$Be as a $\alpha\alpha\Lambda\Lambda$ 
four-cluster system. Accepting the predictive power of model NSC97, 
our calculations suggest that $_{\Lambda\Xi}^{~6}$He 
may be the lightest particle-stable $S=-3$ hypernucleus, 
and the lightest and least strange particle-stable hypernucleus 
in which a $\Xi$ hyperon is bound. Unfortunately, the direct 
production of $\Lambda\Xi$ hypernuclei is beyond 
present experimental capabilities, requiring the use of $\Omega^-$ 
initiated reactions.

\begin{acknowledgments}
This work is partially supported by the trilateral DFG grant GR 243/51-2.
I.N.F. is also partly supported by the Russian Ministry of Education 
grant E00-3.1-133. A.G. acknowledges the support and hospitality of 
the Brookhaven National Laboratory where this manuscript was written up. 
\end{acknowledgments}

\end{document}